\documentclass[11pt]{article}
\def\a{\alpha}

\def\d{\delta}

\def\f{\varphi}

\def\i{\iota}
\def\j{\psi}
\def\k{\kappa}
\def\l{\lambda}

\def\t{\tau}
\def\u{\upsilon}
\def\x{\xi}

\font\cmss=cmss10

\DeclareFontFamily{OT1}{msb}{}{}
\DeclareFontShape{OT1}{msb}{m}{n}
 {  <5> <6> <7> <8> <9> <10> gen * msbm
      <10.95><12><14.4><17.28><20.74><24.88>msbm10}{}
\DeclareMathAlphabet{\bubble}{OT1}{msb}{m}{n}

\def\bR{{\bubble R}}

\chardef\tempcat=\the\catcode`\@
\catcode`\@=11
\def\cyracc{\def\u##1{\if \i##1\accent"24 i
    \else \accent"24 ##1\fi }}
\newfam\cyrfam
\font\tencyr=wncyr10
\def\cyr{\fam\cyrfam\tencyr\cyracc}

\newfont{\goth}{eufm10 scaled \magstep1}
\def\cg{\mbox{\goth g}}

\def\ch{{\cal H}}

\def\cp{{\cal P}}

\def\fr#1#2{{\textstyle{#1\over #2}}}      
\def\p{\partial}

\def\square{\kern1pt\vbox
            {\hrule height 0.6pt\hbox{\vrule width 0.6pt\hskip 3pt
 \vbox{\vskip 6pt}\hskip 3pt\vrule width 0.6pt}\hrule height 0.6pt}\kern1pt}

\def\ra{\rightarrow}

\def\be{\begin{equation}}
\def\ee{\end{equation}}
\def\la#1{\label{#1}} 
\def\re#1{(\ref{#1})} 
\def\arr{\begin{array}{rlll}}
\def\ea{\end{array}}
\def\bea{\begin{eqnarray}}
\def\eea{\end{eqnarray}}

\begin{document}
\rightline{{\cyr TMF} {\bf 123} (2000) 182-188}
\rightline{nlin.SI/0008017}
\vskip 0.5 true cm
\begin{center}
{\Large  Supersymmetric integrable systems  \\[5pt]            
from geodesic flows on superconformal groups}
\footnote{Presented by C.D. at the International Seminar on
Integrable Systems, Bonn, 22nd February, 1999.}
\vskip 0.5 true cm
{\large Chandrashekar Devchand$\;^a$\ \ and\ \ Jeremy Schiff$\;^b$}
\vskip 0.4 true cm
{\it $^a$ Max-Planck-Institut f\"ur Mathematik}\\
{\it Vivatsgasse 7, 53111 Bonn, Germany}\\
{\small devchand@mpim-bonn.mpg.de}\\[3pt]
{\it $^b$ Department of Mathematics and Computer Science }\\
{\it Bar--Ilan University, Ramat Gan 52900, Israel}\\
{\small schiff@math.biu.ac.il}
\end{center}
\vskip 0.5 true cm
\begin{quote}
\centerline{{\bf Abstract}} 
We discuss the possible relationship between geodesic flow, integrability
and supersymmetry, using fermionic extensions of the KdV equation, as well
as the recently introduced supersymmetrisation of the Camassa-Holm equation,
as illustrative examples.
\end{quote}
\vskip 0.5 true cm
\noindent{\bf 1.\ } 
Misha Saveliev was a pioneer of geometric constructions of supersymmetric 
integrable systems (e.g. \cite{ms}). 
In particular, the interrelationship between integrability, geometry, 
diffeomorphism invariance and supersymmetry was a special 
interest of his. We therefore feel that he would have appreciated our recent 
considerations on a) the meaning of integrability for systems containing both 
bosonic and fermionic fields and b) the relation between geodesic flow and
integrability. Although geodesic flows are {\it not} integrable in general, 
many important integrable systems are in fact geodesic flow equations. 
This raises the question of whether one may geometrically determine 
integrable flows amongst geodesic ones. This has in fact been a pressing 
open question for ODEs as well as PDEs ever since Arnold noticed that the
Euler flow equations for incompressible fluids, just like the Euler top 
equations, allow interpretation as geodesic flows on (finite or infinite 
dimensional) Lie groups. Briefly, an inner-product $\langle .,.\rangle$ 
on a Lie algebra $\cg$ determines a right (or a left) invariant metric on 
the corresponding Lie group $G$. The equation of geodesic motion on $G$ 
with respect to this metric is determined by the bilinear operator 
$\;B:\, \cg\times\cg\ra\cg\;$ defined by
\be
\Bigl\langle \bigl[  V, W \bigr] \,,\,   U\Bigr\rangle\ =\ 
\Bigl\langle  W \,,\, B(U,V) \Bigr\rangle\quad , \qquad \forall\ \  W\in\cg\ .
\la{B}\ee
The geodesic flow equation is then simply the Euler equation
\be
U_t = B(U,U)\ .
\la{geodesics}\ee

\noindent{\bf 2.\ }
An important class of examples where geodesic flow is indeed integrable
concerns the group of diffeomorphisms  of the circle, Diff($S^1$). Geodesic
motion  with respect to a metric induced by an $L^2$ norm, $\ \int u^2 dx\ $,
describes Euler flow for a one dimensional compressible fluid, 
$u_t = - 3uu_x\ $, which has implicit general solution, $ut=\fr13 x + F(u)$, 
$F$ an arbitrary function, which describes extremely unstable shock waves. 
If the group is centrally extended to the Bott-Virasoro group, the celebrated 
KdV equation arises, which has extremely stable solutions.
Now, if the metric is changed to one induced by the $H^1$ norm, 
$\ \int (u^2 + \nu u_x^2) dx\ ,\ \nu\in\bR$, one obtains the 
Camassa-Holm (CH) equation,
\be
u_t -  \nu u_{xxt} =  - 3uu_x + \nu(u u_{xxx} + 2 u_x u_{xx})\ .
\la{ch}\ee
This has drawn considerable interest recently as an integrable system 
(having bihamiltonian structure), but displaying more general wave 
phenomena than KdV, such as finite time blow-up of solutions and a class 
of piecewise analytic weak solutions known as {\em peakons}. (For more
complete references we refer to \cite{ds1}).
Nothing is known about what precisely makes these geodesic flows
integrable; the Euler equation for fluid flow in more than one spatial 
dimension is, in general, {\em not} integrable. The investigation of 
families of geodesic flows which include integrable cases may possibly 
yield clues about the special geometric features required. A convenient
way to produce such families, for example containing the KdV and CH 
equations, is to consider geodesic flows on the superconformal group 
containing the Bott-Virasoro group as the even part. This is moreover
a method of generating couplings to fermionic fields: a remarkably rich 
generalisation, as we shall see, of the purely bosonic KdV or CH systems. 
The important question of whether the coupled systems remain integrable has 
hitherto not been adequately explored. 
 
\noindent{\bf 3.\ }
The NSR superconformal algebra consists of
triples $\,\left(u(x),\f(x),a\right)\,$, where $u$ is a bosonic field, 
$\f$ is a fermionic (odd) field and $a$ is a constant. The Lie bracket 
is given by
\be\begin{array}{l}
 \Bigl[ ( u,\f,a)\,,\, ( v,\j,b)\Bigr] 
\la{alg}\\[6pt]
=\  \left( uv_x {-}u_x v{+}\fr12 \f\j\, ,\, 
               u\j_x {-} \fr12 u_x\j {-}\f_x v {+}\fr12\f v_x\, ,\,  
\displaystyle\int_{S^1} dx  
             \bigl(c_1 u_x v_{xx} {+} c_2 u v_x {+} {c_1} \f_x \j_x  
               {+} \fr{c_2}4 \f\j \bigr) \right)\, ,
\ea\ee
where $c_1,c_2$ are constants.
Geodesic flow on the corresponding group with respect to an $L^2$ type 
metric induced by the norm (parametrised by $\a\in\bR$),
\be
\Bigl\langle  ( u,\f,a)\,,\, ( v,\j,b)\Bigr\rangle_{L^2}\
=\ \int_{S^1} dx \left( uv  + \a \f\p_x^{-1} \j  \right)\  + ab 
\la{l2}\ee
yields a  1 parameter fermionic extension of KdV,
\be\arr
u_t  &=& u_{xxx}  - 3 uu_x  + 2  \x\x_{xx}\ ,\\[6pt]
\x_t &=&  \fr1{\a}  \x_{xxx}  - \fr32  u_x \x  - (1+\fr{1}{2\a} ) u \x_x\ ,
\la{ferkdv}\ea\ee
where the fermionic field is defined by 
$\,\f=\l \x_x\ ;\quad \l^2={4\over 3\a}\,$.
In general this family of systems is neither integrable nor supersymmetric
(in the sense of being invariant under supersymmetry transformations 
between $u$ and $\x$, namely $\d u = \t \x_x\ ,\  \d\x = \t u$, where 
$\t$ is an odd parameter). 
Apart from the $\x=0$ KdV case, it is well known that this family 
contains two further integrable cases: 

\noindent
1) $\,\a=\fr14\,$: the kuperKdV system of Kupershmidt, 
which is bihamiltonian, but not supersymmetric,

\noindent
2) $\,\a=1\,$: the superKdV system of Mathieu and Manin-Radul,
which is supersymmetric, but does not afford an extension of
the KdV bihamiltonian structure.

\noindent
It is interesting that both these are particular cases among the 1-parameter 
family of geodesic flows \cite{ds2}. Previously only the bihamiltonian 
kuperKdV system was thought to occur as a geodesic flow \cite{ok}.

Extending \re{l2} to the $H^1$ inner product,
\be
\Bigl\langle  ( u,\f,a)\,,\, ( v,\j,b)\Bigr\rangle_{H^1}\ 
=\ \int_{S^1} dx \left( uv  + \nu u_x v_x 
                      + \a \f\p_x^{-1} \j  + \a\mu \f_x \j \right)\  + ab\ ,
\ee
where $\,\mu,\nu\,$ are further constants, gives rise to the Euler equations
\be\arr
u_t - \nu u_{xxt} 
&=& \k_1 u_x + \k_2 u_{xxx}  - 3 uu_x +  \nu (uu_{xxx}+2u_xu_{xx}) 
        + 2  \x\x_{xx} +  \fr{2\mu}{3} \x_x\x_{xxx}\ ,
\\[6pt]
\x_t - \mu \x_{xxt} 
&=&  \fr{\k_1}{4\a} \x_x +   \fr{\k_2}{\a}  \x_{xxx} 
- \fr32  u_x \x  - (1+\fr{1}{2\a} ) u \x_x
 + \mu u\x_{xxx} + \fr{3\mu}{2}  u_x \x_{xx} + \fr{\nu}{2\a}  u_{xx}\x_x\ .
\la{geo}\ea\ee
Here $\k_1,\k_2$ are independent parameters determined by
$a,c_1,c_2\,$. This is evidently a 5-parameter family of systems containing
CH \re{ch} as well as the 1-parameter KdV family \re{ferkdv}.
It is automatically hamiltonian. Introducing new variables,  
$\,m = u - \nu u_{xx}\,$ and $\,\eta = \x - \mu \x_{xx}\,$, it
may be re-written
\be  \pmatrix{m_t \cr \eta_t } 
   = \cp \pmatrix{ \fr{\d\ch}{\d m} \cr \fr{\d\ch}{\d\eta} } 
\la{hamo}\ee
where the graded hamiltonian structure
\be
\cp =   \pmatrix{ \k_2 \p_x^3 + \k_1 \p_x - \p_x m - m \p_x  & 
                     \fr12\p_x \eta + \eta \p_x \cr
                    -\p_x \eta -\fr12 \eta \p_x &
              \fr3{4\a}(\fr{\k_1}4 + \k_2\p_x^2)- \fr{3}{8\a}m }\ 
\ee 
and the hamiltonian functional is given succinctly in terms of
$U=(u,\f,0)$ by 
\be
\ch_2\ =\ \fr12\ \Bigl\langle   U\,,\, U \Bigr\rangle_{H^1} 
       \ =\  \fr12 \int dx \left( u^2 +\nu u_x^2                   
               +\fr43(\x_x\x + \mu \x_{xx}\x_x) \right)\quad .
\la{second}\ee   
This generalises the so-called {\it second Hamiltonian structure} of  
KdV and its fermionic extensions, as well as that of CH.
The {\it first Hamiltonian structure} of the latter systems does extend 
to systems of the general form \re{geo}, but the single intersection with 
this 5-parameter family is the kuperKdV case. There are therefore no 
bihamiltonian extensions of CH amongst the 
systems \re{geo}. The latter however do contain the superCH system,
\bea
u_t -  u_{xxt} 
&=& - 3 uu_x + 2\x\x_{xx} 
            + uu_{xxx} + 2 u_xu_{xx} + \fr{2}{3} \x_x\x_{xxx}\ , 
\nonumber\\[6pt]
\x_t - \x_{xxt} 
&=&  - \fr32 (u \x)_x + u\x_{xxx} +\fr32 u_x\x_{xx} + \fr12 u_{xx}\x_{x}\ ,
\la{sch}\eea
the unique supersymmetric extension of \re{ch}.
This is invariant under the transformations
$\d u = \t \x_x\ ,\  \d\x =\frac34 \t u$. We have found sufficiently
nontrivial evidence \cite{ds1} to allow us to conjecture the integrability 
of this system.

\noindent{\bf 4.\ }
The meaning of integrability for such systems with fermions remains 
somewhat confusing. Usual arguments for integrability have consisted of 
showing, for instance, the existence of either bihamiltonian structures 
or an infinite number of conserved quantities. Neither of these are very 
reliable criteria. In particular, the superKdV example demonstrates that 
bihamiltonicity is probably sufficient, but by no means necessary for 
integrability of systems with fermions. Discussions of a generalised 
Painlev\'e test for fermionic systems also exist, but the analysis with 
bosonic and fermionic fields taking values in respectively the even and 
odd parts of some grassmann algebra has proven difficult to perform 
carefully enough and there are serious errors in the literature.  
Usually the underlying grassmann algebra is thought of as an infinite 
dimensional algebra, generated by infinitely many odd generators. 
However, integrability should clearly be independent of the choice of 
this underlying grassmann algebra; and choosing some low dimensional 
algebra, generated by a small number of odd generators, 
can yield a great deal of useful information about the general case. 
In particular, consideration of a series of the simplest grassmann algebras 
generated by one, two, three,... odd generators has the power of yielding
conclusive evidence for the non-integrability of the system in general.
With only a finite number of odd generators, the fields afford expansion
in a basis of polynomials of the odd generators. We shall talk of the
{\it n-th deconstruction} when referring to coefficients of up to 
nth-order monomials in the odd generators.
For instance, a bosonic and a fermionic field in the 
{\it second deconstruction} takes values in a grassmann algebra with basis 
$\;\{ 1,\t_1,\t_2,\t_1\t_2\}$, where $\,\t_1\,,\,\t_2\,$ are two odd 
generators. They take the form
\be
u\ =\ u_0\ +\ \t_1\t_2\, u_{1}\ ,\quad  \x\ =\  \t_1 w_1\ +\ \t_2 w_2\ .
\la{d2}\ee
Such {\it deconstructed fields} have purely bosonic components, 
$u_0,u_1,w_1,w_2$,
in terms of which the analysis is considerably more transparent. 
Manton \cite{manton} recently investigated some simple supersymmetric 
classical mechanical systems in this `deconstructive' fashion.

We have investigated the somewhat more general family of 
fermionic extensions of KdV,
\be\arr
u_t &=& -u_{xxx}+6uu_x - \x\x_{xx} \\
\x_t &=& -c\x_{xxx} +a\x u_x + bu\x_x\ ,\la{abc}
\ea\ee
here displayed in rescaled conventions with respect to \re{ferkdv}.
Now, choosing the simplest grassmann algebra, with single odd generator 
$\t$ and basis $\{1,\t\}$, yields the coupling of the KdV field $u$ to 
another bosonic field $w$,
\bea
u_t &=& -u_{xxx}+6uu_x  \la{kdv0}\\
w_t &=& -cw_{xxx} +aw u_x + bu w_x\ ,\la{kdv1}
\eea
where $\x{=}\t w$.
This system, being purely bosonic, allows singularity analysis in a
manner similar to that of Weiss, Tabor and Carnevale. 
The WTC-Painlev\'e analysis for merely this simple system sets strong   
restrictions on the values of $a,b,c$ in the general system \re{abc}
for which integrability remains a possibility.

The WTC algorithm for \re{kdv0}-\re{kdv1} is to seek Laurent series 
solutions in the neighbourhood of an arbitrary singularity manifold, 
$\phi(x,t){=}0$, of the form
\be 
u(x,t)= \sum_{n=0}^\infty  b_n(x,t)\phi(x,t)^{n-2}\ ,\quad
w(x,t)= \sum_{n=n_1}^\infty  a_n(x,t)\phi(x,t)^{n}\ .
\ee
Here the series for $u$ is the standard WTC series for \re{kdv0}, with 
$b_4,b_6$ arbitrary, and the remaining coefficients $b_n$ determined
recursively. The series for $w$ is a Frobenius-Fuchs type series,
with three arbitrary coefficient functions $a_{n_1},a_{n_2},a_{n_3}$
($n_3>n_2>n_1$) and the remainder of the $a_n$'s determined recursively.
It is straightforward to show that $n_1{+}n_2{+}n_3=3$, and as
$N\equiv n_3{-}n_1$ increases, the number of consistency conditions
for existence of the $w$ series increases rapidly. We have examined
all possible cases for $N<10$. For $N{=}8,9$ there are no consistent 
cases, and the list of cases for smaller $N$ is also very limited;
these are tabulated below. It is clear that crucial questions
about the general system \re{abc} may already be decided in the very simple
1st deconstruction. Non-integrability, in particular, seems to make itself
known early on. 

\goodbreak
\begin{center}
\begin{tabular}{|c |c |c  |c |l}
\cline{1-4}
& N &$n_i$ &(a,b,c)&
\\ \cline{1-4}
$i$ & 2 & {0,1,2} & (0,0,c) & uncoupled system
\\ \cline{1-4}
$ii$& 4 &{-1,1,3} &  (3,6,4) & kuperKdV
\\ \cline{1-4}
$iii$& 5 &{-2,2,3} & (3,3,1)&  superKdV
\\ \cline{1-1}\cline{4-4}
$iv$ & &        &(-6,-6,-2) & 
\\ \cline{1-1}\cline{3-4}
$v$ & &{-1,0,4} &  (0,3,1) &  potential form of superKdV
\\ \cline{1-1}\cline{4-4}
$vi$ & &        & (0,-6,-2)&  potential form of $iv$
\\ \cline{1-4}
$vii$&7 &{-3,2,4} & (6,6, 1)& linearisation: $w=\d u$
\\ \cline{1-1}\cline{3-4}
$viii$& &{-2,0,5} & (0,6,1)&  potential form of $vii$
\\
\cline{1-4}
\end{tabular}                                          
\end{center}
\centerline{
{\cmss Table: P-integrable cases within the family \re{kdv1}}}

\noindent
As we see from this list, most of the cases passing the P-test could
have been expected, being 1st deconstructions of kuperKdV ($ii$)
and  superKdV ($iii$), the so-called {\it potential form} of the
latter having solutions given by $w_{(v)}=\int w_{(iii)}\,$, and two trivial
cases: the uncoupled system ($i$) and the linearisation of KdV with 
$w$ denoting the variation $\d u$. There is therefore basically only 
one unexpected case ($iv$), together with its potential form ($vi$). 

Proceeding to the 2nd deconstruction,
with two odd generators and fields having the structure \re{d2},
we note that apart from a multiplicity of $w$'s the only significant
change is the enhancement of the system \re{kdv0},\re{kdv1} by 
an additional equation for $u_1$, the `soul' of $u$,
\be
u_{1t} = -u_{1xxx}+6(u_0 u_1)_x - w_{[1}w_{2]xx}\ ,  
\ee
where $u_0$ satisfies \re{kdv0} and $w_1,w_2$ are solutions of \re{kdv1}.
The P-analysis of this provides a major surprise: {\it $u_1$ has a leading
term of order $\phi^{-3}$}! So coupling fermions to KdV essentially 
changes the singularities of its solutions. In the literature it has 
always been erroneously assumed that the Laurent series
solution leads with a double pole, as for the simple, uncoupled, KdV equation.
This also demonstrates deconstruction to be a very constructive analytical 
tool for the investigation of systems containing fermions. The only cases 
in the above table which do not pass the P-test for the 2nd deconstruction
are the unexpected cases $iv$ and $vi$. In general, therefore, these cases
do not correspond to new integrable fermionic extensions of KdV. The 1st
deconstructions are, however, certainly P-integrable. This illustrates a
further important advantage of this deconstructive method: It provides
a simple routine for the construction of integrable couplings of systems
known to be integrable. 

\noindent{\bf 5.\ }
With manifold choices available for the underlying grassmann algebra,
both superCH \re{sch} and the integrable cases amongst \re{abc} provide
a very rich class of solvable systems. In fact they include some important
classical ODEs as reductions. For instance, in the galilean reference frame,
with all fields depending on only one variable 
$z{=}x{-}vt$,  CH (just as KdV)  corresponds to the familiar equation,
\be
p'^2 = 1 - 2c_1 p +c_2 p^2 - \fr2v p^3\ ,
\la{w0}\ee
where $c_1, c_2$ are integration constants. This equation has a well-known
general solution in terms of the Weierstra\ss\ $\wp$-function,
\be
p(z) = - 2v \wp(z) + \fr16 c_2 v  \ ,
\la{wp}\ee
where the periods of $\wp$ are determined by the coefficients $c_1,c_2,v$.
In this galilean reference frame, the first and second deconstructions 
of super/kuper KdV, for which \re{w0} is the `body', are special cases of
the Lam\'e equation which have been of interest for over a century.
For instance, for the superKdV case, the first deconstruction yields 
the equation
\be
w(z)'' - \left(6\wp(z-z_0) - \fr{v}{2} \right) w(z) = d\ ,\quad d=const.
\ee

In the case of our new superCH system, we have shown in \cite{ds1} that the
1st deconstruction and the galilean reduction of the 2nd deconstruction fulfil
the  requirements of the P-test. These are therefore nontrivial integrable
reductions of the superCH system \re{sch}. In the process we have encountered 
further integrable generalisations of Lam\'e's equation which seem to be new: 
In the 1st deconstruction the fermionic equation has galilean reduction,
\be
h'{}' 
+\fr38\left({p\over v}-{c_2\over 6}+{c_1\over p} -{7\over 2p^2}
\right)h\ =\ 0\ ,
\la{lame1}\ee
and the soul of the bosonic equation in the 2nd deconstruction yields
\be
k'{}' + \left({3p\over v}-{c_2\over 4}-{3\over 4p^2} \right) k\ =\  0\ , 
\la{lame2}\ee
where, in both equations, $p$ is given by the linear expression in $\wp$ \re{wp}.
To conclude we note that integrability for PDEs is certainly not an 
all-or-nothing affair. Even if superCH turns out to be non-integrable in general, 
we have shown that it does display nontrivial integrability properties and it
would be interesting to understand the geometry underlying our analytical results.

\def\Large{\large}

\end{document}